\documentclass{aa}  

\usepackage{graphicx}
\usepackage{amsmath}
\usepackage{caption}
\usepackage{subcaption}
\DeclareMathOperator*{\argmin}{argmin}
\usepackage{txfonts}
\newcommand{\be}{\begin{equation}}
\newcommand{\ee}{\end{equation}}

\begin{document}

   \title{Combining summary statistics with simulation-based inference for the 21 cm signal from the Epoch of Reionization}
   \titlerunning{}

   \author{B. Semelin\inst{1}\fnmsep\thanks{benoit.semelin@obspm.fr}
           \and
           R. Mériot\inst{1}
           \and
           A. Mishra \inst{1}
           \and
           D. Cornu \inst{1}
         }

   \institute{Sorbonne Université, Observatoire de Paris, PSL university, CNRS, LERMA, F-75014 Paris, France\\
            }

   \date{Received X X, 2022; accepted X X, 2022}

 
\abstract{
The 21 cm signal from the Epoch of Reionization will be observed with the up-coming Square Kilometer Array (SKA). SKA should yield a full tomography of the signal which opens the possibility to explore its non-Gaussian properties. How can we extract the maximum information from the tomography and derive the tightest constraint on the signal? In this work, instead of looking for the most informative summary statistics, we investigate how to combine the information from two sets of summary statistics using simulation-based inference. To this purpose, we train Neural Density Estimators (NDE) to fit the implicit likelihood of our model, the LICORICE code, using the Loreli II database. We train three different NDEs: one to perform Bayesian inference on the power spectrum, one to do it on the linear moments of the Pixel Distribution Function (PDF) and one to work with the combination of the two. We perform $\sim 900$ inferences at different points in our parameter space and use them to assess both the validity of our posteriors with Simulation-based Calibration (SBC) and the typical gain obtained by combining summary statistics. We find that our posteriors are biased by no more than $\sim 20 \%$ of their standard deviation and under-confident by no more than $\sim 15 \%$. Then, we establish that combining summary statistics produces a contraction of the 4-D volume of the posterior (derived from the generalized variance) in $91.5 \%$ of our cases, and in $70$ to $80 \%$ of the cases for the marginalized 1-D posteriors. The median volume variation is a contraction of a factor of a few for the 4D posteriors and a contraction of 20 to 30 $\%$ in the case of the marginalized 1D posteriors. This shows that our approach is a possible alternative to looking for \emph{sufficient statistics} in the theoretical sense. }

\keywords{Cosmology: dark ages, reionization, first stars --  Radiative transfer-- Early Universe -- Methods: numerical }
\titlerunning{Combining 21-cm signal statistics for inference}
\maketitle
%

\section{Introduction}

The 21 cm signal emitted by the neutral hydrogen in the intergalactic medium during the Cosmic Dawn (CD) and the Epoch of Reionization (EoR) is one of the most promising probes of the early universe. Indeed, its spatial fluctuations and time evolution encode information on both the astrophysics of the first galaxies and on cosmology. A number of observational programs are underway to detect the signal. The EDGES experiment presented a possible detection of the global (i.e. sky-averaged) signal during the CD \citep{Bowman18}. However, the SARAS-3 experiment excluded a EDGES-like signal with 95 $\%$ confidence \citep{Singh22}. Radio-interferometers such as LOFAR, MWA, HERA or NenuFAR provide much more information to work with than global signal experiments when attempting to separate the cosmological signal from the foreground emission. One can either avoid the foregrounds by selecting a region in Fourier space where they are weak, or model and subtract them carefully. This has led to establishing upper limits on the power spectrum of the signal \citep[e.g.][for recent results]{Yoshiura21, Abdurashidova22, Munshi24, Acharya24}. Some of these, \citet{Abdurashidova22} and \citet{Acharya24}, are low enough to disfavor some scenarios of reionisation. With the upcoming Square Kilometer Array (SKA), these upper limits will hopefully turn into a detection over a range of wavenumbers and redshifts, allowing to constrain the models tightly.

With the prospect of an actual detection in the next few years, we can turn our attention to the best way to extract information from the observations. Given a parametric model able to predict the signal, the most straightforward method is to perform Bayesian inference on the power spectrum with a Markov Chain Monte Carlo (MCMC) algorithm \citep[e.g.][]{Greig15}. This approach is strongly motivated by the fact that the power spectrum will likely be the first summary statistics of the fluctuations of the signal detected with high confidence. On a longer time scale however, we can hope to obtain a full tomography of the signal. This is meaningful because it is well known that the 21 cm signal is non-Gaussian and thus the power spectrum does not contain all the information. A number of alternative and/or higher order summary statistics have been explored to characterize the 21 cm signal. The bi-spectrum has been studied, for example, by \citet{Shimabukuro16, Majumdar18, Watkinson17, Hutter20} and similar three-point statistics by \citet{Gorce19}. The Pixel Distribution Function as been investigated by \citet{Ciardi03,Mellema06, Harker09, Ichikawa10, Semelin17} and summary statistics based on wavelets by \citet{Greig22,Hothi24, Zhao24}. All these summaries contain information about the astrophysics of the 21 cm signal that can be estimated with a Fisher analysis \citep{Greig22, Hothi24} and quantified in detail with Bayesian inference \citep{Zhao22, Zhao24, Prelogovic23}. The main difficulty is that the likelihoods of the observed values of these summaries for given model parameters may well not be Gaussian. Even if we assume them to be Gaussian as an approximation, we do not have an analytic expression for the means and correlation matrix like we do (under some approximations) in the case of the power spectrum. This is where simulation-based inference (SBI hereafter) comes into play \citep[e.g.][]{Papamakarios16, Alsing19}.  

Indeed, SBI does not require an explicit analytical form for the likelihood. It relies on a (large) number of parameter-observable pairs generated with the model and, considering them as samplings of the likelihood implicit to the model, fits an approximate parametric form of the likelihood to this collection of samplings. The fit is high-dimensional and is commonly entrusted to a neural networks. This procedure as been applied to the 21 cm signal by \citet{Zhao22, Prelogovic23, Zhao24, Meriot24b}. The advantage of SBI is twofold, first it allows to use any (combination of) summary statistics of the observed signal without any theoretical formula for the likelihood and second, it allows the use of more complex and costly models. Indeed, as found for example in \citet{Meriot24} the typical size of the learning sample required to train the neural network is typically two orders of magnitude smaller than the number of steps in a classical MCMC chain for a 4-5 dimensional parameter space and these samples can be generated independently of each other from the prior rather than sequentially. This has made it possible to perform Bayesian inference for a model that consists in performing a full 3D radiative hydrodynamics simulation \citep[e.g][]{Meriot24}.

The possibility to perform Bayesian inference on any summary statistics opens  questions on how to choose those statistics. Formally, $n$ quantities can be \textit{sufficient statistics} and deliver optimal inference (in terms of Fisher information) in an  $n$-dimensional parameter space. Methods to optimally compress the signal and obtain sufficient statistics have been proposed both in cases where the likelihood is known \citep{Alsing18} and in cases where it is not \citep[e.g][]{Charnock18}. The later requires training a neural network whose loss attempts to maximize the Fisher information, but training such a network requires a large enough training sample that may be difficult to generate in real applications. As an alternative, different summary statistic can be compared systematically to assess which ones provide the most information \citep[e.g.][]{Greig22,Hothi24}. In this work we explore another option: to use SBI to combine summary statistics and thus increase the amount of information used in the inference. Combining observables for inference usually involves assumptions about their independence that are not necessarily valid. This is even more the case when combining summary statistics of the same observable that are most likely correlated through instrumental noise. However, SBI is fully able to account for these correlations, the limit being the expressivity of the chosen parametric form for the likelihood. In this work, we will show how SBI can be used on radiative hydrodynamics simulations to combine the power spectrum and the pixel distribution function of the $21$ cm signal to obtain tighter constrains than with either of these individually.

In section 2, we briefly present our model (the LICORICE simulation code), our training sample (the Loreli II database) and the summary statistics we will be using. In section 3 we recall the principle of SBI, describe our implementation and present validation tests using simulation-based calibration. In section 4 we assess statistically the gains obtained on the posteriors by combining statistics. Finally, in section 5, we present our conclusions.

\section{Model, data and statistics}

\subsection{The model: 3D radiative transfer with LICORICE and SPINTER}

Bayesian inference has traditionally relied on algorithms such as MCMC that require a large number of forward passes with the model (from model parameters $\theta_i$ to simulated data $d_i$). A typical number is $10^5$ to $10^6$ passes for a 4 to 5 parameter inference. Moreover, these passes have to be performed sequentially, although the EMCEE algorithm \citep{Goodman10} alleviates this requirement to some extend. The model we use in this work is a 3D radiative hydrodynamics code, LICORICE, to compute the local density, temperature and ionization fraction, combined with the SPINTER code to compute the local Lyman-$\alpha$ flux. While these quantities allows to compute the 21-cm signal, each pass from parameters to signal requires hundreds of CPU hours even at moderate resolution. Consequently, as far a inference is concerned, such a model can only be used to build a database (see \ref{LoReLi}) that serves as a training sample for  machine learning approaches: training an emulator or various flavours of SBI.

The LICORICE code allows to model the 21-cm signal. It is a particle-based code that computes gravity with an Oct-Tree algorithm \citep{Barnes86}, and hydrodynamics with the Smooth Particle Hydrodynamics algorithm \citep[e.g.][]{Monaghan92}. Validation tests are provided in \citep{Semelin02}. Monte Carlo Ray-Tracing transfer was added, both for the UV to compute the ionization of Hydrogen and Helium \citep{Baek09} and for the X-ray to compute the heating of neutral hydrogen by secondary electrons \citep{Baek10}. Later on, coupling between radiative transfer and hydrodynamics was implemented, along with an hybrid MPI+OpenMP parallelization \citep{Semelin16}. While this allowed to leverage large computing infrastructures to run massive simulations, the total computing cost of a single large simulation is in the $10^6$ CPU-hour range, which is incompatible with building a large enough database of simulations to span an interesting parameter range. This is the reason why \citet{Meriot24} implemented several improvements that enable moderate resolution simulations to produce realistic 21 cm signals. An out-of-the-box $256^3$ particle simulation in a $\sim 300$ cMpc box resolves only halos with masses larger than $\sim 10^{11}$ M$_\odot$: these appear late and produce barely any ionization in the IGM by $z\sim 6$. The contribution from halos with masses below the simulation resolution was added relying  on the Conditional Mass Function formalism \citep{Lacey93,Rubino08}. \citet{Meriot24} showed that this approach allowed a $256^3$ simulation to produce a star formation history similar to that of a $2048^3$ simulation. Another improvement was to implement the option to treat the gas content in a particle with two distinct phases in terms of temperature. Indeed, at low resolution, the typical size of an SPH particle is (much) larger than the ionization front thickness, and a partial ionization actually represents a juxtaposition of cold neutral gas and warm ionized gas. Assigning the same temperature to both phases biases the 21 cm signal, in particular by weakening the signal in absorption that requires cold gas. The LICORICE code outputs provide the density, temperature and ionisation state of hydrogen necessary to compute the 21 cm signal.

The SPINTER code \citep{Semelin23} is used  in post-treatment to compute the local Lyman-$\alpha$ flux and thus the last ingredient for the 21 cm signal: the hydrogen spin temperature. The SPINTER code include the effect of scatterings in the wings of the Lyman-$\alpha$ line that change the profile of the flux around a source \citep[e.g.][]{Semelin07,Chuzhoy07, Reis21, Semelin23}. It does not however account for the effect of the bulk velocities that can also have an impact \citep{Semelin23, Mittal23}.

\subsection{The Data: Loreli II database}
\label{LoReLi}
The Loreli II database is described in details in \citep{Meriot24b} and summary statistics of the simulations are publicly available \footnote{https://21ssd.obspm.fr}. We will here present only briefly its main features. The simulations are run in a $200$ h$^{-1}$cMpc box, with $256^3$ particles and fixed cosmological parameters. Each simulation has a different realisation of the initial density field and is run down to a redshift when the IGM is fully ionized (typically between $z=5$ and $z=7$), saving $\sim 50$ snapshots along the way. Each particle snapshot is gridded, then the local Lyman-$\alpha$ flux is computed with SPINTER and $21$-cm signal cubes are produced. The $\sim 9800$ simulations correspond to different combinations of the following parameters \citep[see][for detailed definitions and values]{Meriot24b} :

\begin{itemize}
    \item $f_X \in [0.1,10]$ the efficiency of X-ray emissivity,
    \item $\tau \in [7.5,105]$ Gyr the time scale for conversion of gas into stars in star forming dark-matter halos,
    \item $M_{min} \in [10^8, 4.\times 10^9]$ M$_\odot$, the minimum mass for a halo to be allowed to form stars,
    \item $r_{H/S} \in [0,1]$ a ratio of the energy emitted per second in hard X-ray (e.g. from X-ray binaries) to the energy in both hard and soft X-ray (e.g. from Active Galactic Nuclei) 
    \item $f_{esc} \in [0.05, 0.5]$ the fraction of photons escaping from galaxies. This accounts for unresolved recombination in high density regions around the sources.
\end{itemize}
Note that the sampled volume is not an hypercube, in particuliar $(\tau,M_{min})$ pairs have been selected so that simulated Luminosity Functions are a reasonable match to observed ones.

When generating the summary statistics on which simulation-based inference is performed, an important step is to properly account for stochastic processes in the modelling. In particular, when several types of summary statistics are combined, they should be computed from the same realisation of the stochastic processes to account for possible correlations. In our case, this happens naturally for the first source of stochasticity: cosmic variance. Indeed, each point in the parameter space is explored with a single simulation, and thus a single realisation of the cosmic density field. The second source of stochasticity that our model takes into account is the thermal noise of the instrument. We generate thermal noise realisations directly in visibility-space following \cite{McQuinn06}, before transforming back to image-space. We generate 1000 realisations, and obtain 1000 coeval noise cubes at each redshift that can be added to the signal cubes. We choose an instrumental resolution of $6$ arcmin by applying a gaussian filter in visibility space, and we implement a 30-$\lambda$ cutoff. This resolution allows a $10$ mK noise level on the tomography with the SKA observation setup described below. Then, the signal and noise are summed and the resulting noised cubes are used to compute the different summary statistics, thus preserving the possible correlations due to the thermal noise. In this work, the setup used for computing the thermal noise is the layout with 512 stations with a diameter of $35$ m corresponding to the 2016 SKA baseline design\footnote{https://www.skao.int/sites/default/files/documents/d18-SKA-TEL-SKO-0000422\_02\_SKA1\_LowConfigurationCoordinates-1.pdf}, filled with 256 dipole antenna with min$(2.56, \lambda^2 )$ m$^2$ effective collecting area and system temperature of $T_{sys}=100+300\left({\nu \over 150 \mathrm{MHz}}\right)^{-2.55} K$. The current baseline design differs slightly (e.g $38$ m diameter for the station), but this has little consequence for the conclusions of this work. Finally, we considered $100$ hours of observations in sessions of 8 hours centered on the zenith. Indeed, as argued in \citet{Meriot24b}, the parameter space sampling of the Loreli II database may be too sparse to confidently estimate the narrow posteriors resulting from a signal perfectly evaluated from $1000$ hours of observation.

\subsection{Summary statistics}
The summary statistics that we will use in the inference section are computed from coeval cubes of the 21 cm signal. While using slabs from the lightcone would be more realistic, it would raise questions on how the different statistics and their combination respond to a non-stationary, non-isotropic signal, which is beyond the scope of this study. However, in computing the posteriors, we do include the information from the evolution of the signal by considering the joint likelihood  of the statistics computed at 3 different redshifts. We used $z=8.18, 10.32$ and $12.06$. From Fig. 3 in \citet{Meriot24b}, we can see that these values span the most significant range for the evolution of the signal in Loreli II. Note however that including a lower-redshift value may improve the weak constraints on $f_{esc}$ that we find.
\subsubsection{Power spectrum}
The first summary statistics that we consider is computed from the standard isotropic power spectrum $P(k)$, defined through the relation:

\be
\langle \delta T_b(\mathbf{k}),  \delta T_b(\mathbf{k^\prime)} \rangle = P(k)(2 \pi)^3 \delta_D(\mathbf{k}+\mathbf{k^\prime})
\ee
\noindent
where $\mathrm{k}$ and $\mathrm{k^\prime}$ are Fourier modes, $\delta T_b$ is the signal and $\delta_D$ the Dirac distribution function. The $\langle \rangle$ notation is formally an ensemble average. Computing the isotropic spectrum ($P$ depends only on $|\mathbf{k}|$), allows us to replace this ensemble average by an average over Fourier modes falling in the same spherical shell. In practice we use the so-called dimensionless power spectrum $\Delta^2(k)={k^3\over 2 \pi^2} P(k)$, that estimates the contribution of logarithmic bins to the total power in the signal. Since we use a $6$ arcmin resolution when generating the noise cubes, corresponding to a $\sim 12$ h$^{-1}$cMpc scale at these redshifts, we first downsample our $21$ cm cubes from $256^3$ to $32^3$, that is $\sim 3$ arcmin per pixel. Then we compute the power spectrum of the noised cubes using non-overlapping logarithmic bins with a step of $\sqrt{2}$ between bin boundaries, resulting in 8 bins between wavenumbers $0.03$ h cMpc$^{-1}$ and $0.5$ h cMpc$^{-1}$. For the inference we discard the two lowest wavenumber bins that are the most affected by cosmic variance and the highest wavenumber bin that is the most affected by thermal noise. This leaves us with 15 summary statistics: $5$ power spectrum bins at 3 different redshifts. Note that we infer on the biased power spectrum. We do not subtract the average power of the noise. This ensures that the summary statistics are positive and makes for easier normalisation of the data when training the Neural Density Estimators.

\begin{figure*}
    \centering
    \includegraphics[width=\textwidth]{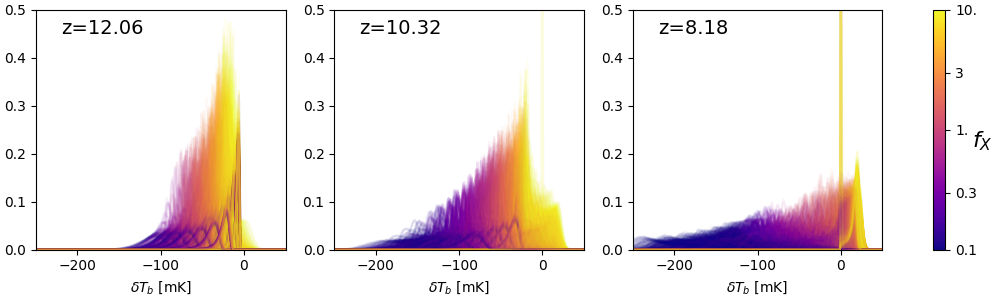}
    \caption{Pixel distribution functions of all the simulations in the Loreli II database, illustrating the variety of possible shapes. The color of each individual curve depends on the  $f_X$ value for the corresponding simulation.}
    \label{fig:loreli_pdfs}
\end{figure*}

\subsubsection{Pixel Distribution Function}
The Pixel Distribution Function (hereafter PDF) is a simple one-point statistics that nevertheless encodes non-gaussian information, unlike the power spectrum. It simply counts the number of pixels in the signal that falls in each bin of intensity values. Fig. \ref{fig:loreli_pdfs} gives a sense of the variety of shapes of the PDFs in the Loreli 2 database. Our goal is to summarize the information in the PDF with a number of statistics similar to what is used for the power spectrum, that is 5. Let us first note that the most informative statistics of the PDF is the mean because it tells us whether the signal is in absorption or emission at a given redshift, which is strongly connected to the model parameter values. However, the mean of the signal is not available in interferometric observations. Thus, before computing the PDF, we subtract the mean from each slice of the cubes perpendicular to an arbitrarily chosen line-of-sight. Thus we infer on shifted PDFs that all have zero mean. The first idea would to define a small number (e.g. 5) of bins and use their population as summary statistics. This is actually much less straightforward than it seems. The PDFs of various cubes can have very different width: choosing narrows bins results in ignoring information outside the bin range for some cubes, while choosing wide bins may result for some cubes in all the pixels falling in the same bin. Also, having bins with a 0 population for a large fraction of the model parameter values result in singular behaviours for the likelihood that are tricky to handle. Moreover, the distribution of bins population in response to thermal noise is non trivial: near empty bins may exhibit Poisson-like distributions, while highly populated bins may be gaussian, and since thermal noise displaces pixels from one bin to another, the resulting noised populations are obviously correlated, making for a more complex likelihood. Our attempts at using the bin values as summary statistics did not yield satisfactory results.

Another possibility is to use the statistical moments of the PDF computed from the cubes as: $m_k={1 \over N_{pix}} \sum_{i=1}^{N_{pix}} \delta T_b(x_i)^k$, where $N_{pix}$ is the number of pixels and $x_i$ their positions. Note that these moments are centered because we subtract the mean of the signal from the cubes. One possible difficulty with using these summary statistics is that, if we want 5 of them, from $m_2$ to $m_6$, we will raise the noised $\delta T_b$ value to powers up to 6, making our statistics sensitive to outliers. An alternative to the $m_k$ are the linear moments \citep{Hosking90}, defined as follows. Considering $n$ independent samples from the distribution, we note $X_{r:n}$ the r-th smallest value in the samples. Then the linear moment $l_k$ is defined as:

\be
l_k={1 \over k} \sum_{r=0}^{k-1}(-1)^r \binom{k-1}{r} \mathbb{E}(X_{k-r:k})
\ee
\noindent
One can check that, for example, $l_2$ is proportional to the expectation of the largest minus expectation of the smallest in samples of 2 draws: thus it is a measure of the width of the distribution. In the same manner, upper linear moments estimate properties of the distribution similar to those estimated by traditional statistical moments, but without resorting to raising the sample to powers larger than 1. They are thus expected to be less sensitive to outliers and to yield lower variance estimators. Generally speaking, when used as targets for the inference, lower variance statistics are expected to produce more constraining posteriors on the model parameters. As computing the $l_k$ is significantly more expensive than the $m_k$, we use the Fortran routines provided by \citet{Hosking96}, that we find to be hundreds of times faster than the available Python version. We will use moments $l_2$ to $l_6$, evaluate their performances for inference compared to moments $m_2$ to $m_6$, and compare and combine them with the power spectrum. 

Note that the statistical moments and the power spectrum are not fully independent summary statistics: for example $m_2$ is the variance of the signal and thus very much correlated with the sum of the power spectrum. Indeed, our goal is not to maximize information with a minimal number of summary statistics but to demonstrate the possibility to combine them with SBI and reach tighter constraints on the model parameters.

\section{Simulation-based inference with Neural density estimators}

\begin{figure*}[ht]
    \centering
    \includegraphics[width=\textwidth]{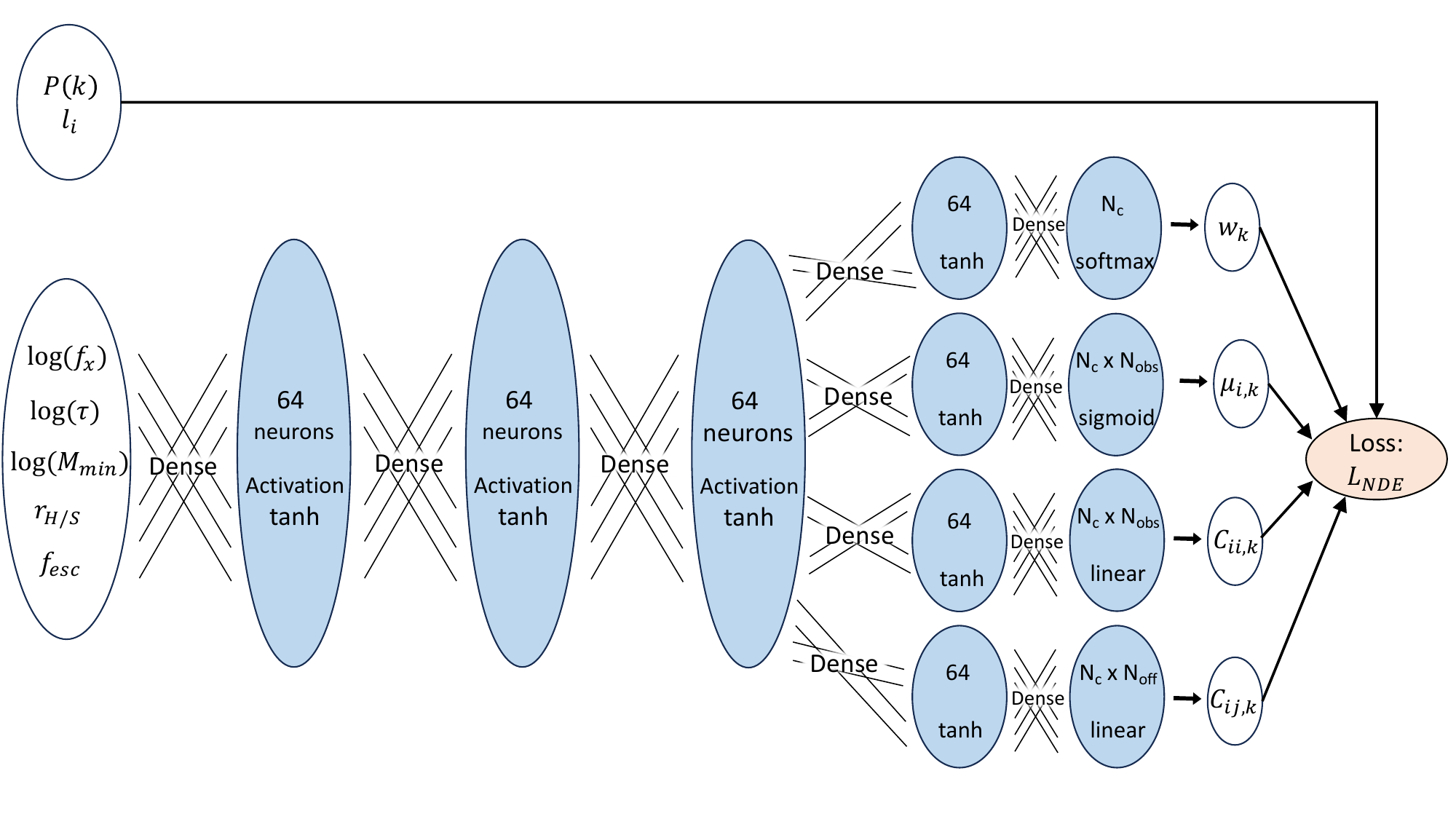}
    \caption{Architecture of the neural network used for the NDEs. The inputs are the model parameters and the summary statistics of a corresponding noised signal and the outputs allow to evaluate the parametric likelihood function of the summary statistics (see main text). For each layer (blue regions) we give the number of neurons and the activation function. $N_c$ is the number of components of the Gaussian Mixture Model describing the likelihood, $N_{obs}$ is the length of the vector of summary statistics, and $N_{off}={(N_{obs}+1)N_{obs} \over 2}$.}
    \label{fig:NDE_diag}
\end{figure*}

\subsection{Principle}
The principle of SBI is presented for example in \citet{Alsing19}. The fundamental idea is that a complete model, including stochastic processes such as cosmic variance or instrumental noise, is able to simulate some observed data $d$ (a noised 21 cm cube in our case) from the set of parameters $\theta$ (those described in section 2.2 in our case). Then, each pass forward of the model produces a pair $(\theta,d)$ which is a sampling of the underlying joint distribution of $\theta$ and $d$, or given $\theta$, a sampling of the likelihood $P(d|\theta)$. In many cases, we do not have an analytic expression of the true likelihood, as for example for the PDF moments. However, given a large number of such samples, one can attempt to fit a parametric function to the true likelihood. As the parameters of the fitting function will depend, in the general case, on $\theta$, the fitting procedure will require many samples for each $\theta$, and even a simple multivariate Gaussian parametric likelihood will involve $(n+2)(n+1)/2$ fitting parameters in the case of a vector of data with $n$ components. In our case with vectors of $15$ or $30$ components, this means fitting $120$ or $496$ parameters. Consequently the task of finding the best fit is entrusted to a Neural Density Estimator (NDE). The NDE (often a simple Multilayer Perceptron) will take as input the $(\theta,d)$ pairs and produce the $\theta$-dependent values of the parameters of the parameterized likelihood function. $d$ will only be used during training to converge to the correct fitting parameters values. Let us note $L_{True}(d|\theta)$ the true, unknown likelihood, and $L_{NDE}(d|\theta,\alpha_i)$ the parametric likelihood that can be estimated from the output of the NDE. The $\alpha_i$ are the outputs of the NDE and functions of $\theta$ and of the NDE weight. Then, training the NDE means minimizing the difference between $L_{True}$ and $L_{NDE}$. This difference can be quantified with the Kullback-Liebler divergence:

\be
KL(L_{True},L_{NDE})=\int L_{True}(\mathbf{d}|\theta) \log\left( L_{True}(\mathbf{d}|\theta) ) \over L_{NDE} (\mathbf{d}|\theta,\alpha_i) \right) d\mathbf{d}
\ee

The evaluation of the integral can be replaced by a summation on the available samples: the probability that each sample was generated is accounted for by the $L_{True}(\mathbf{d}|\theta)$ factor before the logarithm in the integral. Then, since we are working to minimize the KL with respect to the $\alpha_i$, the numerator term in the logarithm, that does not depend on the $\alpha_i$, can be ignored, and our goal is to find:
\be
\argmin_{\alpha_i} KL(L_{True},L_{NDE})= \argmin_{\alpha_i}\sum_{j}^{N}-\log L_{NDE} (\mathbf{d_j}|\theta_j,\alpha_i)
\ee
This quantity can be readily computed for a set of $(\theta_i,\mathbf{d}_i)$ and used as a loss for the NDE, that will adjust its weights to produce the $\alpha_i$ that minimize this loss.

\subsection{Inference framework}

\subsubsection{General setup}
The first and possibly most important design element when building the NDE framework is the choice of the parametric form for the likelihood function. The most basic choice is a multivariate Gaussian. Indeed, our observables are summary statistics that are typically combinations of many physical quantities affected by stochastic processes. From the central limit theorem we generally expect them to exhibit near-Gaussian distributions. However, we are not necessarily able to compute analytically the parameters of these distributions (e.g for the PDF moments), and even less their full covariance matrix when the summary statistics are correlated. This is why SBI is meaningful even with a simple multivariate Gaussian parameterization. However, to better account for departures from the Gaussian behaviour, we can use Gaussian Mixture Models or Masked Autoregressive Flows or MAFs \citep{Alsing19, Prelogovic23}. However, \citet{Prelogovic23} did not find a clear improvement from using MAFs to perform inference on the power spectrum. Moreover, \citet{Meriot24b} find that, with the Loreli II database and thermal noise resulting from 100 hours of SKA observations, a simple multivariate Gaussian likelihood for the power spectrum performs best. Since, in this work, we do infer on the PDF that is able to encode non-Gaussian behaviours, the result presented here are all obtained with a 2-component Gaussian Mixture Model.  The observed lack of improvement beyond 2 components did not motivate us to explore MAFs.

We will do the inference for the transformed parameters: $\log_{10}(f_x)$, $\log_{10}(\tau)$, $r_{H/S}$, $\log_{10}(M_{min})$ and $f_{esc}$. This is consistent with a regular sampling in the Loreli II database. We will report our findings in terms of these quantities normalized to the $[0,1]$ interval. The sampled 5-dimensional volume serves as our prior (see in \citet{Meriot24b} how it differs from an hypercube in the $(\tau,M_{min})$ plane). The prior is flat inside the sampled volume and zero outside. 
As is usual with neural networks we benefit much from normalizing all the data, so we also need to normalize the observable quantities. For the (noised) power spectrum, we first apply the $P(k,z) \longrightarrow \log_{10}(1+P(k,z))$ transformation, then, individually for each of the 15 different $(k_i,z_j)$ pairs used for the inference, we normalize to the $[0,1]$ interval using the minimum and maximum values found across the whole range of noised signal in the database. We do the same for the $l_k$ moments of the PDF, without the $\log(1+x)$ transformation. The same normalization does not work for the $m_k$ moments who are much more sensitive to outliers. In this case we first compute the mean and variance of the distribution of $l_k$ across the database and do a linear transformation to bring the values from the $[\mu-2\sigma,\mu+2\sigma]$ interval to the $[0,1]$ interval.

\subsubsection{Designing and training the NDE}

With our choice for the form of the parametric likelihood, the NDE is trained to minimize the following loss:

\be
Loss=-\log\left[\sum_{k=1}^{N_c} {w_k \over \sqrt{(2 \pi)^{N_{obs}} \mathrm{det}(\Sigma})} e^{-{{1 \over 2}(\mathbf{d}-\boldsymbol{\mu})^T\Sigma^{-1}(\mathbf{d}-\boldsymbol{\mu})}}\right]
\ee
where $N_c$ is the number of components in the mixture model, $N_{obs}$ the dimension of the vector of summary statistics, $w_k$ the weights of the components as predicted by the NDE, $\boldsymbol{\mu}$ the vector of the means of the summary statistics predicted by the NDE, and $\Sigma$ the covariance matrix of the summary statistics. The NDE does not predict directly the covariance matrix, but the coefficients of the triangular matrix $C$ that is related to the inverse of the covariance matrix by the Cholesky decomposition: $\Sigma^{-1}=CC^T$. Predicting $C$ automatically ensures that $\Sigma$ is definite-positive. In the general case, the $w_k$, $\boldsymbol{\mu}$ and $\Sigma$ are functions of the input parameters. 

The architecture of the NDE is described in Fig. \ref{fig:NDE_diag}, and implemented with the Keras-tensorflow framework. In our configuration, using a deeper network or increasing the number of neurons in the layers did not results in a decreased validation loss. The training sample is made of $90\%$ of the simulations in Loreli II selected at random. For each simulation, $1000$ different realisations of the instrumental noise are added to the 21 cm cubes and the summary statistics are computed from the noised cubes, resulting  $\sim 8\times 10^6$ samples. The $10$ remaining percent of simulations are used for validation. The Nadam optimizer is used with an initial learning rate of $0.0002$. A small batch size of 4 samples gives a lower loss while further reducing the batch size is prohibitive in term of efficiency. We first train the networks by providing the $\sim 8\times 10^6$ samples once, then we decrease the learning rate by a factor of 10 and retrain on half of the samples to improve convergence. The loss function does not decrease any more beyond that point. 

Using the final value of the validation loss is a usual measure of the quality of the training of the network, however, in our case, this must be handled with care. Indeed, given the form of our loss function, its value depends on the choice of normalization of the summary statistics, that changes the covariance matrix $\Sigma$. Changing the dimension of the summary statistics vector also changes the final loss value. Moreover we do not have a known lower bound of the loss like, for example, when using a mean squared error loss. Thus, while in some cases we can tell that one network performs better than another, we cannot tell from the loss value how well it performs on an absolute scale. We have to turn to a much more costly evaluation of the performances: simulation-based calibration (SBC).

\subsubsection{MCMC sampler}
Once the NDEs are trained they can be used in a classical MCMC framework to compute the likelihood of the target signal at each step of the chain. Even if the cost of each likelihood computation is just a few milliseconds, computing $10^5-10^6$ steps sequentially represents a non-negligible cost for a single inference. To speed up the process we use the EMCEE sampler \citep{Goodman10} that involved creating a team of 160 walkers that can be advanced simultaneously in parameter space . Moreover we used a custom implementation of EMCEE rather than a standard Python package. Indeed, when the likelihood function involves a forward pass in a neural network, it is important to use rather large batches of inputs for good compute performance. Passing a likelihood function that involves a neural network to the usual Python implementation of EMCEE results in batches of size 1 and can be up to 30 times slower than our custom implementation where the batch size is half the number of walkers.

\subsection{Validating the framework with Simulation Based Calibration}

\begin{figure*}
    \centering
    \includegraphics[width=\textwidth]{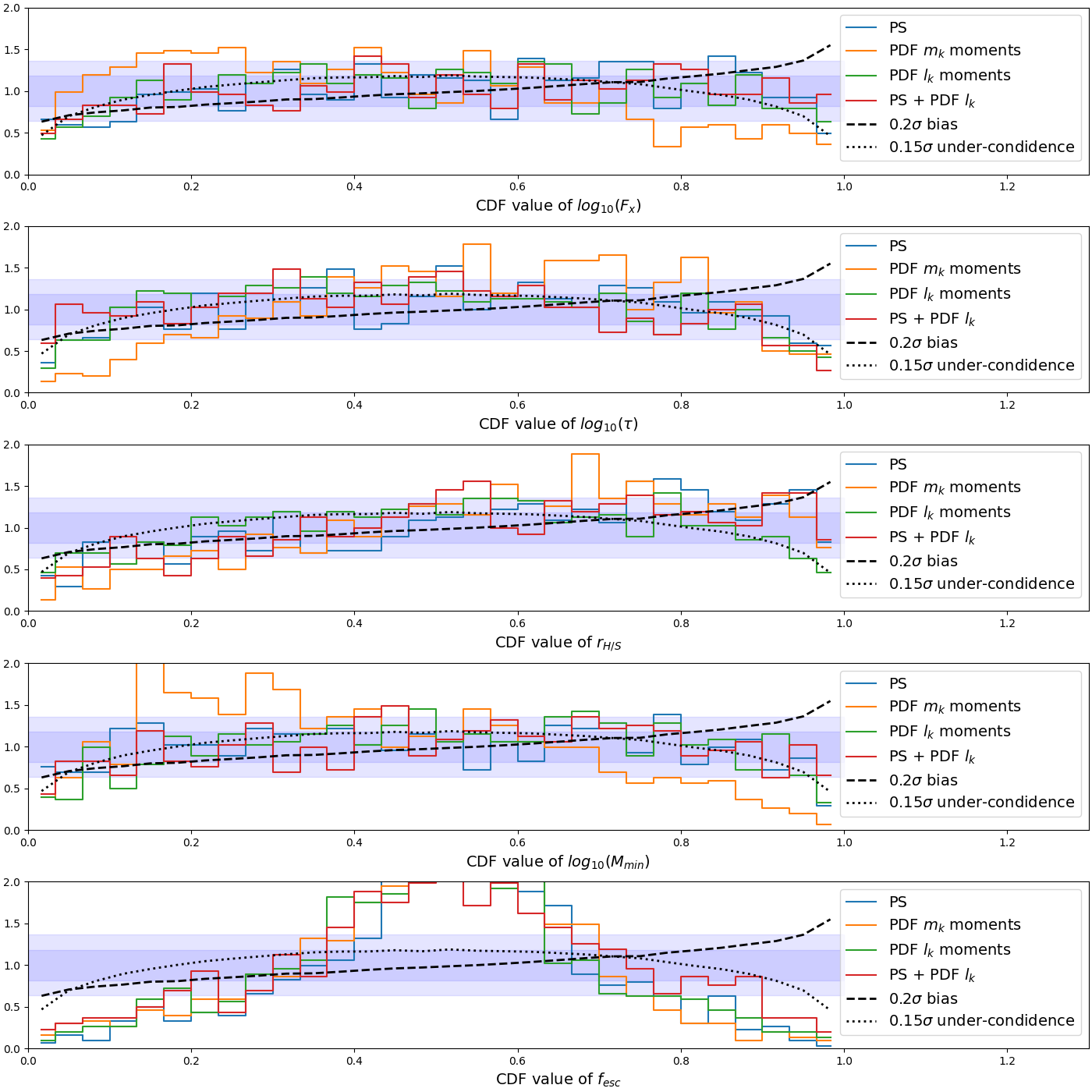}
    \caption{Histograms over the 908 inferences describing where the true parameter values fall in the cumulative distribution function of the 1D marginalized posteriors. One histogram is plotted for each different NDE. If the posteriors match the true unknown posteriors the histogram should be flat. A toy model is plotted for comparison where the approximate posterior is a biased or under-confident Normal distribution and the true posterior is the Normal distribution.}
    \label{fig:sbc}
\end{figure*}

\begin{figure*}[ht]
    \centering
    \begin{subfigure}{0.49\textwidth}
        \centering
        \includegraphics[width=\linewidth]{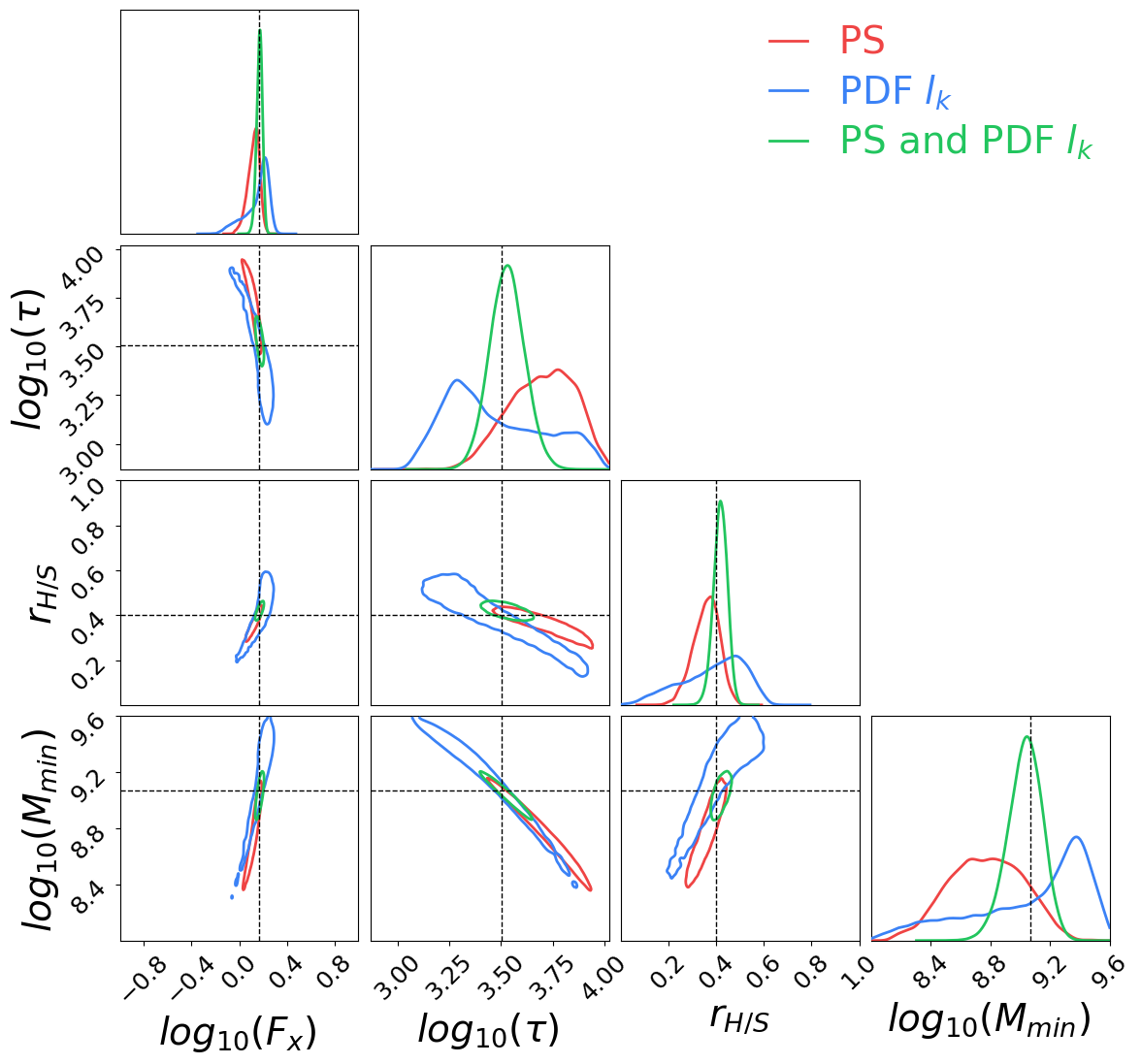} 
        \caption{} \label{fig:corner_a}
    \end{subfigure}
    \begin{subfigure}{0.49\textwidth}
        \centering
        \includegraphics[width=\linewidth]{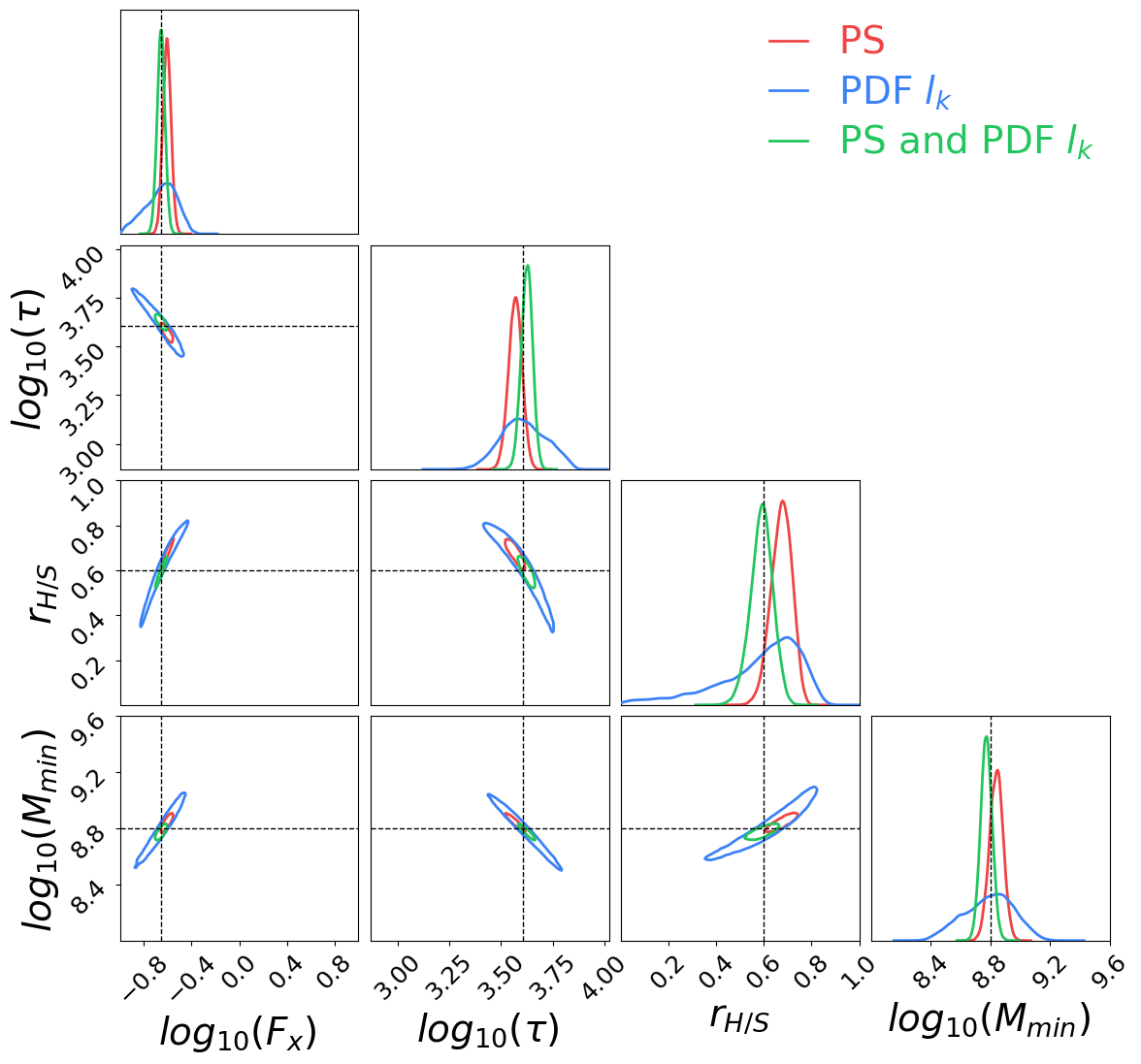} 
        \caption{} \label{fig:corner_b}
    \end{subfigure}

    \vspace{0.5cm}
    \begin{subfigure}{0.49\textwidth}
        \centering
        \includegraphics[width=\linewidth]{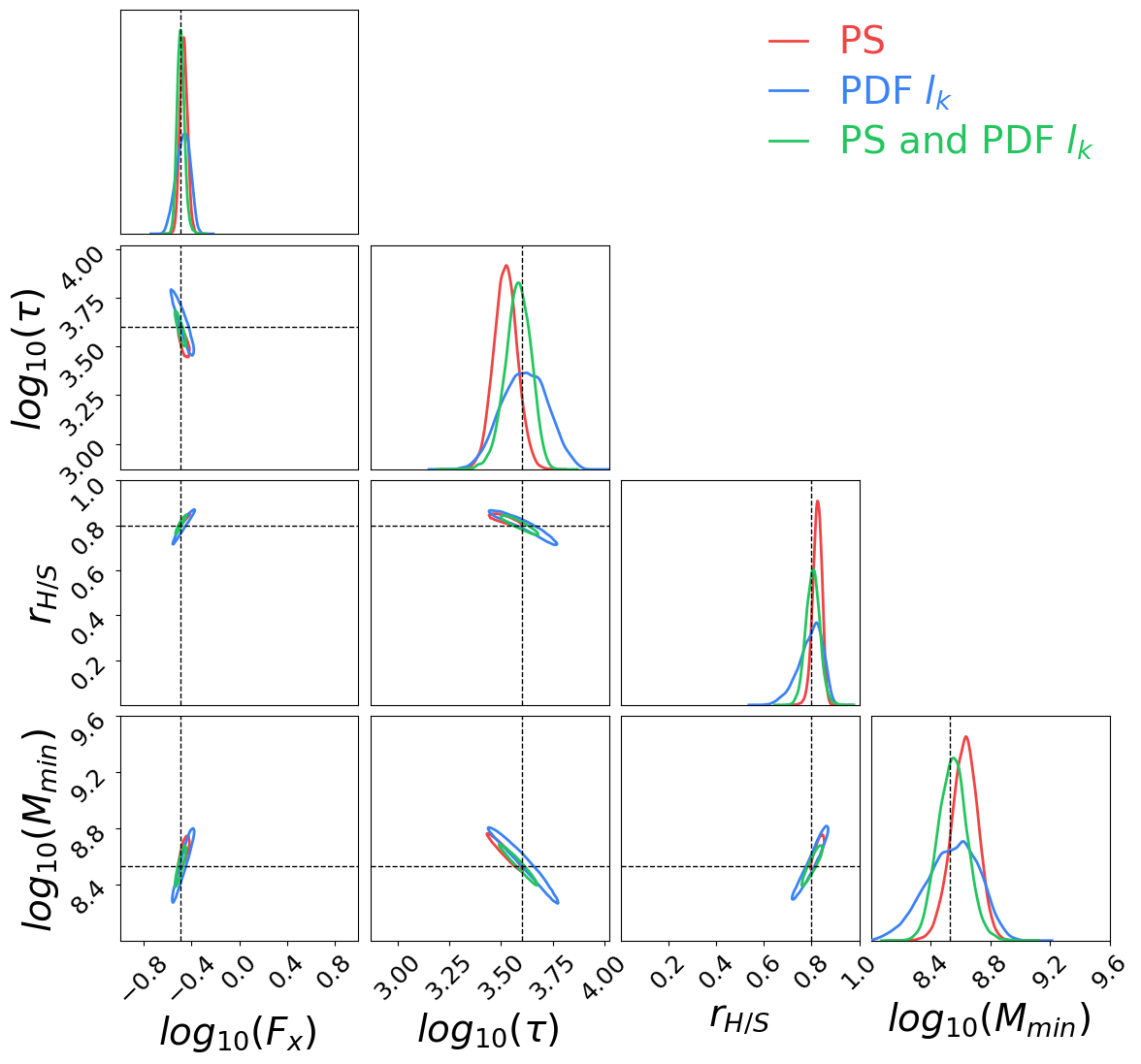} 
        \caption{} \label{fig:corner_c}
    \end{subfigure}
    \hfill
    \begin{subfigure}{0.49\textwidth}
        \centering
        \includegraphics[width=\linewidth]{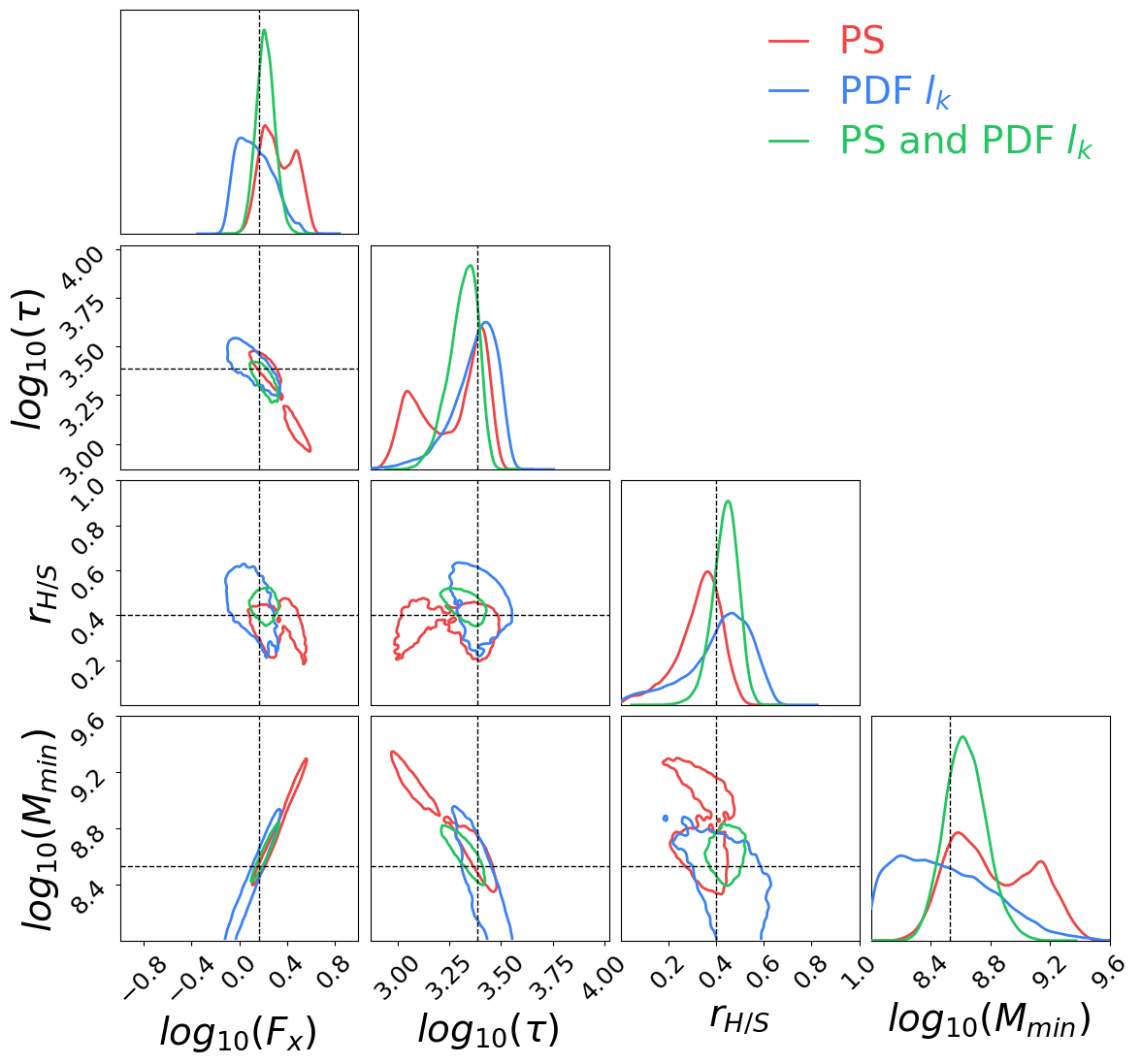} 
        \caption{} \label{fig:corner_d}
    \end{subfigure}
    
    \caption{Corner plot of the inferences done with the 3 NDEs on the different summary statistics for 4 different signals (one in each panel). The corresponding truth value of the parameters are indicated by the black dashed lines. Only the 1-sigma contour are plotted for clarity.}
    \label{fig:corner}
\end{figure*}

The difficulty when using Machine Learning to perform Bayesian inference is to evaluate the validity of the resulting posteriors. The methods can be tested on toy-models where the posterior is known analytically but there is no guaranty that satisfactory convergence will extend to cases where the model is more complex. SBC is a partial answer to the validation issue. It is presented in \citet{Talts18} and applied to the 21-cm signal inference in \citet{Prelogovic23} and \citet{Meriot24b}: it validates the shapes of the marginalized 1D posteriors (note however that two different posterior in higher dimension can still produce identical 1D maginalizations). 

The principle relies on the same fact as SBI: a $(\mathbf{d},\boldsymbol{\theta})$ pair resulting from a forward pass of the model (e.g. a Loreli II simulation together with a noise realization) is a draw from the true joint distribution implicit to the model. Then, $\boldsymbol{\theta}$ is a draw from the \emph{true} posterior $P_{True}(\boldsymbol{\theta}|\mathbf{d})$. Consequently each component $\theta^i$ of $\boldsymbol{\theta}$ will fall with equal probability in any quantile of the marginalized 1-D posterior $P_{True}(\theta^i|\mathbf{d})$. A necessary condition for the inference framework to be valid is that this property also holds for the posterior computed with the NDE. As this test is statistical in nature, it requires a large number of inferences on different $\mathbf{d}_k$. To find the quantile of the 1-D maginalized posterior in which each of the  components $\theta^i_k$ falls, one can consider the MCMC chain computed during the inference (after removing the burn-in phase), compute the rank statistics  of $\theta^i_k$ among all the values of $\theta^i$ in the chain and directly deduce the quantile or, equivalently, the corresponding value in the cumulative distribution function (CDF) of the marginalized posterior. After analysing a large number of inferences in such a way, one can build an histogram of the resulting population of the quantiles and this histogram should be flat.

We estimated the validity of the inferences performed with 4 different NDEs respectively trained for inference on the power spectrum, the $m_k$ PDF moments, the $l_k$ PDF moments, and the combination of power spectrum and $l_k$ moments. As learning near the sharp edges of Loreli II prior in parameter space is a more difficult task for the NDE, we selected 908 simulations not located near the boundary of the prior, performed the inference with all four NDEs and computed the quantiles histograms. The results are presented in fig. \ref{fig:sbc}. The first obvious conclusion is that the inference on $f_{esc}$ is strongly under-confident: too many true $f_{esc}$ values fall near the median of the posterior (0.5 value of the $x$ axis). This unsatisfactory result is understandable since only $3$ different values of $f_{esc}$ are sampled in Loreli II. Since we excluded the boundary values, all the simulations used for the SBC have the same $f_{esc}$. We are obviously in a regime where our sampling of the parameter space is not dense enough to run an SBC test comfortably. The second conclusion is that the NDE trained to infer on the $m_k$ moments does not performs as well as the others. In particular the marginalized posteriors for $\log_{10}(F_x)$ and $\log_{10}(M_{min})$ are more biased as revealed by the tilted histograms: the true value of the parameter falls more often on one side of the median of the computed posterior. A possible explanation is that the higher sensitivity of the $m_k$ moments to outliers makes it more difficult for the NDE to learn the likelihood.

The other cases (power spectrum, $l_k$ moments, and combination of the two) show similar, rather satisfactory behaviours for parameters other than $f_{esc}$. For comparison, the same diagnostic is computed in the cases where the estimated posterior is a Normal distribution with mean $0$ and standard deviation $1$ and the true posterior is either a Normal distribution with mean $0.2$ and standard deviation $1$ (biased estimated posterior), or a Normal distribution with mean $0.$ and standard deviation $0.85$ (under-confident estimated posterior). We can see on the figure that a bias equal to $\sim 20 \%$ of the variance or a variance overestimated by $\sim 15\%$ are reasonable estimations of the level of inaccuracy in the estimated marginalized posteriors. The higher dimensional data used by the NDE dealing with the combination of power spectrum and PDF moments does not seem to degrade the performances. Considering the uncertainties and approximations in the modelling of the 21-cm signal even with full radiative hydrodynamics simulations \citep[see e.g.][]{Semelin23}, and the level of uncorrected systematics that currently plague the observations and will probably not completely disappear with the SKA, we consider the obtained level of accuracy satisfactory. Of course, if a thermal noise level consistent with $1000$ hours of observation was used, the same inference framework may not yield sufficiently accurate posteriors.

\section{Comparing and combining summary statistics}

Since we found that the inference on $f_{esc}$ is not quite reliable, we will from now on only reports results on the other 4 parameters. Note that $f_{esc}$ is still varied in the MCMC chains. Also, since the the NDE training with the $m_k$ moments performs the worst in terms of SBC, we will mostly focus on the $l_k$ moments.

\subsection{Individual posteriors}
For the computation of the SBC diagnostics we performed $908$ inferences on noised signals corresponding to different points in the parameter space. The resulting posteriors show a variety of behaviours and we cannot show them all here. Before we present a statistical analysis of the results, however, we can see in Fig. \ref{fig:corner} examples of the result of the inference for 4 different signals. Case (a) shows a situation where inferring on the combination of power spectrum and PDF linear moments brings a significant tightening of the constraints on all parameters compared to the inference on the power spectrum or the PDF linear moments separately. Case (b) is similar but with a less significant tightening of the constraints compared to the power spectrum. These behaviours are quite common in the 908 inferences as shown below in the statistical analysis. Case (c) shows an unusual case where constraints using the power spectrum alone are actually tighter than constraints using the combination of power spectrum and PDF moments. This counter-intuitive situation may happen for example when adding the PDF moments shifts the median of the posterior towards a region of the parameter space where the signal intrinsically provides weaker constraints on the parameters. Alternatively, this may be a sign that the training of the NDEs are locally less accurate. Finally case (d) shows a situation where the power spectrum posterior is bimodal and adding the information contained in the PDF moments lifts the ambiguity.

\subsection{Statistical analysis}

\begin{figure}
    \centering
    \includegraphics[width=\columnwidth]{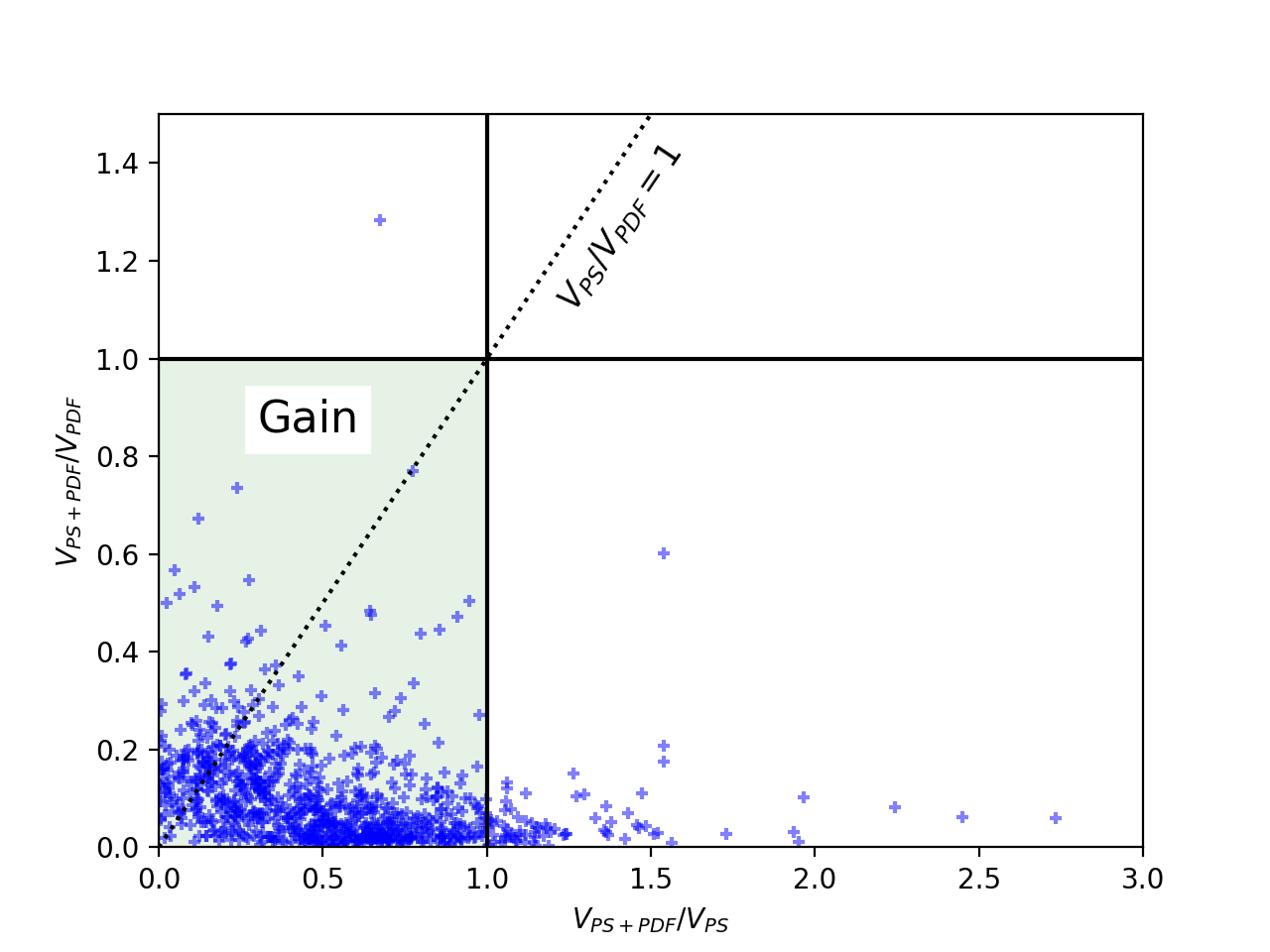}
    \caption{Scatter plot of the ratios of the characteristic volume occupied by the 4D posteriors (see main text for definition). Each blue cross is an inference on a different noised signal. In the green region the combined statistics give a smaller volume than either of the individual statistics. }
    \label{fig:Gen-var-scatter}
\end{figure}

\renewcommand*\arraystretch{1.1}
\begin{table} 
\centering                    
\begin{tabular}{c c c c c}                
Statitics & PS + PDF $l_k$ & PS & PDF $l_k$ & PDF $m_k$ \\
\hline                                 
    $\langle V \rangle \times 10^7$ & 0.48 & 1.33 & 7.53 & 8.58 \\
    $\langle\sigma_{\log_{10}(f_X)}\rangle$& 0.019 & 0.024 & 0.0375 & 0.0369 \\
    $\langle\sigma_{\log_{10}(\tau)}\rangle$ & 0.052 & 0.075 & 0.080 & 0.087 \\
    $\langle\sigma_{r_{H/S}}\rangle$ & 0.035 & 0.049      & 0.060 & 0.067 \\
    $\langle\sigma_{\log_{10}(M_{min})}\rangle$ & 0.063 & 0.087     & 0.109 & 0.109 \\
\hline                                           
\end{tabular}
\caption{Median values of the square root of the generalized variance (V) of the 4D posterior and the standard deviations of the 1D marginalized posteriors for inference on different summary statistics. The average are computed on the 908 inferences performed at different locations in parameter space. }              
\label{table:1} 
\end{table}

\begin{figure*}
    \centering
    \includegraphics[width=\textwidth]{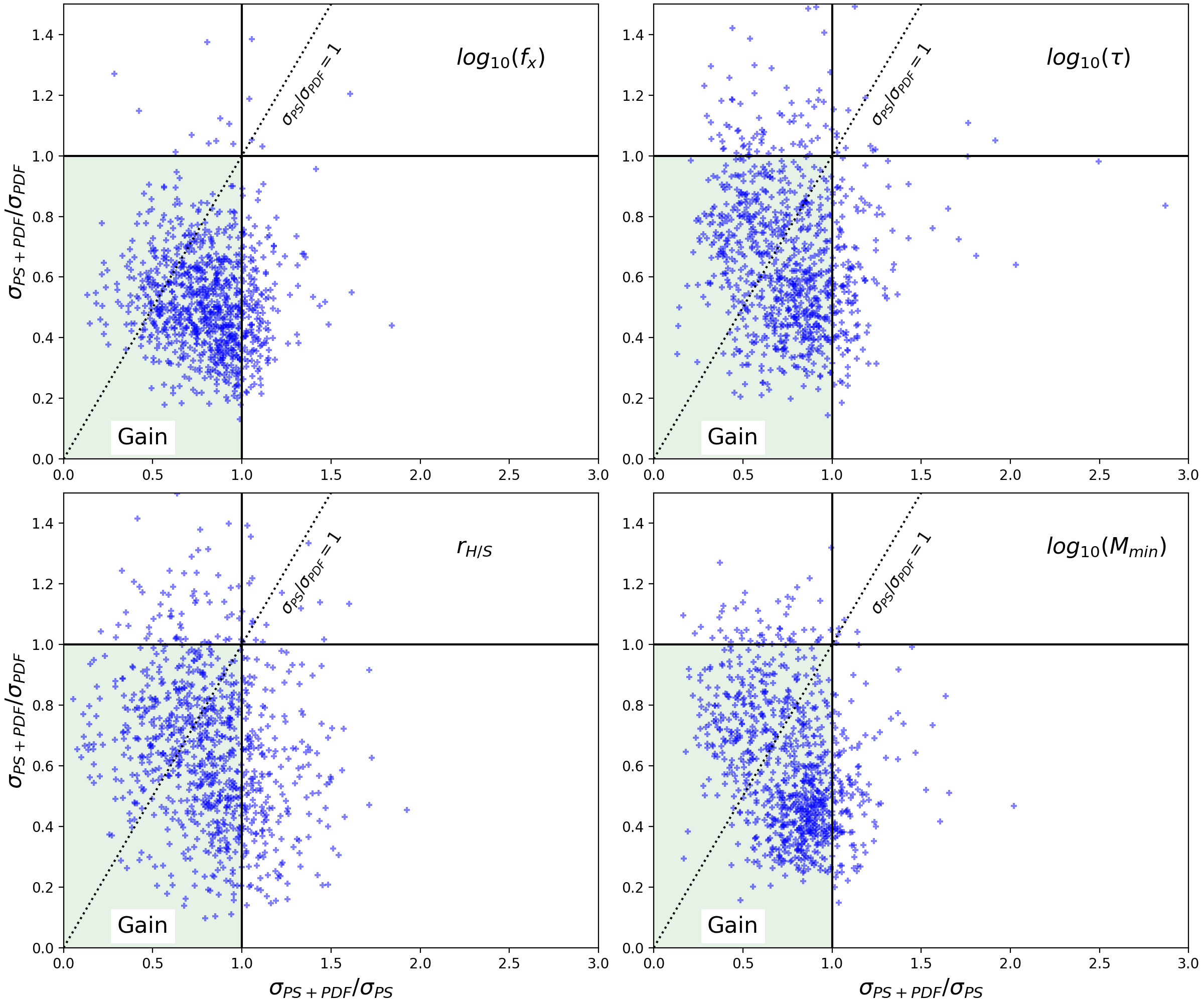}
    \caption{Same as fig. \ref{fig:Gen-var-scatter} but for the ratios of the volumes (i.e. standard deviation here) occupied by the 1D marginalized posteriors.}
    \label{fig:Param-var-scatter}
\end{figure*}

The question we want to answer is: does using the combination of power spectrum and linear moments of the PDF gives, statistically, tighter constraints on the parameters than using either alone. In the 4-dimensional parameter space (excluding $f_{esc}$ from the analysis), the tightness of the constraints can be quantified by the square root of the so-called generalized variance: $V=\det(\Sigma)^{1 \over 2}$, where $\Sigma$ is the covariance matrix of the posterior that we can easily estimate from the MCMC chains. If the posterior is a multivariate Gaussian, $V$ is proportional to the volume enclosed in the $1\sigma$ hyper-surface. In more general cases it still characterizes the volume encompassed by the posterior. The other obvious quantities that we can use are the standard deviations of the marginalized 1D posteriors. In table \ref{table:1} we report the median of all these quantities computed from the posteriors at the $908$ different location in parameter space  and using the different possible summary statistics as signals. The first and main conclusion is that, in terms of the median values of the standard deviations, using the combination of the power spectrum and the linear moments of the PDF brings a tightening of constraints for all parameters of $20$ to $30 \%$ and a factor more than $2.5$ for the volume $V$ of the 4-D posterior. The second result is that the power spectrum produces tighter constraint than the PDF. Finally, the $m_k$ moments of the PDF perform slightly worse than the $l_k$ moments (but remember that the posteriors for the $m_k$ are also less reliable).

We have seen in fig \ref{fig:corner}, however, that in some cases, inference on the power spectrum gives tighter constraints than on the combined statistics. To estimate if this is a common occurrence, Fig. \ref{fig:Gen-var-scatter} shows a scatter plot of $V_{PS+PDF} \over V_{PDF}$  against $V_{PS+PDF} \over V_{PS}$ for the 908 inferences. A dot below both the $x=1$ and $y=1$ lines indicates that the combined statistics outperforms both of the individual statistics. We can see that a small but non-negligible fraction (computed to be $8.5 \%$) of the points in parameter space actually gives tighter constraints from an individual statistics, almost always the power spectrum. The figure also shows that the PDF alone does outperform the power spectrum alone (points above the $y=x$ line) in $15 \%$ of the cases. Fig. \ref{fig:Param-var-scatter} shows the same plots but for the standard deviation of the 1-D marginalized posteriors for each of the parameters. The fractions of cases were a single statistics outperform the combination is $19 \%$, $25 \%$, $33 \%$ and $21 \%$ for $\log_{10}(f_x)$, $\log_{10}(\tau)$, $r_{H/S}$, and $\log_{10}(M_{min})$ respectively. While these numbers are higher than for $V$ (accounting for cases where only one or two parameters out of 4 are better constrained by the power spectrum or PDF alone), the overall result is that combining summary statistics produces a significant improvement on the constraints on the parameters by exploiting more information, a result that can only be achieved in our case using SBI.

\section{Conclusion}

In summary we have used the Loreli II database of radiative hydrodynamics simulations of the 21 cm signal to perform simulation-based inference using Neural Density Estimators to model the likelihood with a 2-component Gaussian Mixture Model. We considered the following summary statistics as targets: the power spectrum, the statistical moments of the PDF, the linear moments of the PDF, and the combination of power spectrum and linear moments of the PDF. Using simulation-based calibration,  we showed that in most cases the 1-D marginalized posteriors are biased by no more than $20 \%$ of the variance and under-confident by no more than $15 \%$. However, marginalized posteriors on $f_{esc}$ were strongly under-confident because of the inadequate sampling of this parameter in Loreli II. Also, posteriors using the statistical moments of the PDF where less accurate than the others. Then we showed that combining summary statistics by inferring simultaneously on the power spectrum and linear moments of the PDF brings a tightening of the $1\sigma$ volume of the 4-D posterior in $90 \%$ of the cases. For 1D marginalized posteriors, this tightening occurs in 70-80 $\%$ of the cases and has a median amplitude of $20$ to $30 \%$.

The limitation of Loreli II for use in SBI are discussed in \citet{Meriot24b}. Learning from Loreli I, the sampling steps were designed to match, on average, the standard deviation of the posteriors for $100$ hours of SKA observation. A lower level of thermal noise may require a denser sampling, and the more peaked likelihood function may not be as well fitted with a simple Gaussian Mixture Model, requiring the use of more expressive neural network architectures such as MAFs. Another limitation worth mentioning is that only the cosmic variance and the instrumental thermal noise were considered as sources of stochasticity. A more realistic model would also include stochasticity in the source formation process sourced at scales not resolved in the simulations and by the chaotic nature of the dynamics of galaxy formation and evolution on long time scales. These limitations come down to a simple fact that we must always keep in mind: the  result of an inference may be valid given the underlying model, but its usefulness is still limited by the validity of the model itself.


\begin{acknowledgements}
The Loreli II database used in this work was build using computer and storage resources provided by GENCI at
TGCC thanks to the grants 2022-A0130413759 and 2023-A0150413759 on the supercomputer Joliot Curie's ROME partition. 
\end{acknowledgements}

\bibliographystyle{aa}
\bibliography{myref}

\end{document}